\documentclass{aa}  

\usepackage{graphicx}
\usepackage{txfonts}
\usepackage{natbib}
\usepackage{hyperref}
\bibpunct{(}{)}{;}{a}{}{,} 

\usepackage{color}

\newcommand\Fixed[1]{}

\newcommand\HII{H\textsc{ii}}

\newcommand{\kpc}{{\rm kpc}}
\newcommand{\magnitude}{{\rm mag}}
\renewcommand{\deg}{^{\circ}}
\newcommand\loneq[1]{\ensuremath{\ell = #1\deg}}

\newcommand\deq[1]{\ensuremath{D= #1~\kpc}}
\newcommand\dbe[2]{\ensuremath{#1 \leq D \leq #2~\kpc}}
\renewcommand\ao{A_0}
\newcommand\prob[1]{P\left(#1\right)}
\newcommand\probhat[1]{\hat{P}\left(#1\right)}
\newcommand\probind[2]{P_{#2}\left(#1\right)}
\newcommand\probc[2]{P\left(#1 \mid #2\right)}
\newcommand\probchat[2]{\hat{P}\left(#1 \mid #2\right)}
\newcommand\probcind[3]{P_{#3}\left(#1 \mid #2\right)}
\newcommand\gbp{G_{BP}}
\newcommand\grp{G_{RP}}
\newcommand\FoG{Finger-of-God}

\begin{document}

\title{FEDReD II : 3D Extinction Map with 2MASS and Gaia DR2 data}


\author{
  C. Hottier \inst{1} 
  \and 
  C. Babusiaux \inst{2,1}         
  \and
  F. Arenou \inst{1}
}

\institute{
  GEPI, Observatoire de Paris, PSL University, CNRS ; 5 Place Jules Janssen 92190 Meudon, France
  \and 
  Univ. Grenoble Alpes, CNRS, IPAG, 38000 Grenoble, France
}

\date{Received 27 January 2020 ; accepted 18 June 2020}

\abstract
{}
{We aim to map the 3D distribution of the interstellar extinction of the Milky Way disk up to  distances larger than those probed with the Gaia parallax alone. } 
{We apply the FEDReD (Field Extinction-Distance Relation Deconvolver) algorithm to the 2MASS near-infrared photometry together with the Gaia DR2 astrometry and photometry. 
This algorithm uses a Bayesian deconvolution approach, based on an empirical HR-diagram representative of the local thin disk, in order to map the extinction as a function of distance of various fields of view. }
{We analysed more than 5.6 million stars to obtain an extinction map of the entire Galactic disk
  within $|b|<0.24\deg$. This map provides information up to $5~\kpc$ in the direction of the
  Galactic centre and at more than $7~\kpc$ in the direction of the anticentre. This map reveals the
  complete shape of structures known locally, such as the Vela complex or the split of the local
  arm. Furthermore our extinction map shows many large ``clean bubbles'' especially one in the
  Sagittarius -- Carina complex, and four others which define a structure that we nickname the
butterfly.}
{}


   \maketitle
%

\section{Introduction}
The interstellar extinction attenuates the light coming from background objects. Moreover
this attenuation is a function of the wavelength of radiation, which causes
the reddening phenomenon.  The study of extinction is mandatory to recover the
absolute magnitude and intrinsic colours of Galactic or extragalactic stars. 

Mapping the extinction also provides access to the distribution of dust. This component of the Galactic disk
is itself an interesting key to understand the evolution of the Milky Way. In fact, high dust
density areas are expected to be associated with high star formation regions. 
Mapping the extinction of the Galactic disk is thus
a way to study the spatial structure of spiral arms.

In the past decades, several methods have been developed in order to draw 3D extinction maps of the Galactic disk.
With limited resolution and distances, the first results were obtained 
by \citet{1968AJ.....73..983F, 1980A&AS...42..251N, 1992A&A...258..104A,1997AJ....114.2043H}.
\cite{marshall2006} compared the 2MASS photometry to the stellar population
synthesis of the Besan\c{c}on Model \citep{robin2012} to obtain extinction density. This approach
was also used by \cite{chen2013} and \cite{schultheis2014} which added Glimpse and VVV catalogue
to their dataset. \cite{sale2014} applied a
hierarchical Bayesian method to infer stellar properties and extinction profile with the IPHAS
survey. Also using a Bayesian method, \cite{green2014} derived an extinction map from Pan-STARRS data,
later combined with 2MASS \citep{green2015,green2018} and Gaia DR2 \citep{green2019} data.
\cite{capitanio2017, lallement2019} used a Bayesian inversion technique described in
\cite{vergely2001} on composite dust proxies with Gaia DR1 parallax and on 2MASS crosmatched with Gaia DR2 respectively.
\cite{rezaeikh.2018} developed a non-parametric 3D inversion on APOGEE data to obtain density map of
the solar neighbourhood.
\cite{chen2019a} used a random forest algorithm on the crossmatch of Gaia DR2, 2MASS and WISE, trained on spectroscopic data, to
infer the local extinction density.

In this work we use the Field Extinction-Distance Relation Deconvolver (FEDReD) algorithm described 
in \cite{babusiaux2020}.
This is a Bayesian deconvolution algorithm which uses an empirical H-R diagram to study photometry and parallax, 
taking into account the completeness of the field of view under study, to derive both the extinction as 
a function of distance and the stellar distance distribution. 
We apply the algorithm to data of the Two Micron All Sky Survey
\citep[2MASS,][]{skrutskie2006} crossmatched with the second Data Release of Gaia \citep[DR2,][]{gaiacollaboration2018a}. 
This paper is organised as follows : in
Section~\ref{sec:data} we present the data and filters used to select stars.
Section~\ref{sec:fedred} is a quick sum up of the FEDReD method. In Section~\ref{sec:merging}
we detail our method to merge every field of view results into a self-consistent
map. Section~\ref{sec:results} presents our extinction map, with the details of the different
visible features and their relation with the other components of the Galactic disk. 

\section{Data}
\label{sec:data}
This work uses photometry and astrometry based on two surveys, Gaia and 2MASS.
Gaia DR2 data provides high precision parallax
and photometry in $G$, $\gbp$ and $\grp$ bands
2MASS provides photometry in three near - infrared bands: $J$, $H$ and
$K_s$. As Gaia DR2 and 2MASS are full sky surveys, we are able to draw an extinction map of the
entire Galactic disk.

The 2MASS catalogue is used as a basis for our analysis, as its completeness is easier to model
than the Gaia one, the latter being more dependent on the crowding as well as on the Gaia's
scanning law.  Every star that we study has therefore 2MASS photometry which can be completed
with Gaia-DR2 parallax, photometry or both.

To select stars in the 2MASS catalogue, we used the \texttt{ph\_qual}
flag and kept every star with at least a \texttt{ph\_qual}=\texttt{D} for each photometric
band. 

We use the Gaia-2MASS crossmatch provided by \cite{marrese2019}. When a 2MASS source had more
than one Gaia best neighbour, we did not associate any Gaia information to the 2MASS source.

For the Gaia photometry \citep{evans2018}, we do not use $\gbp$ and $\grp$ photometry for stars
affected by crowding issues using the filter \texttt{phot\_bp\_rp\_excess\_factor} $> 1.3+0.06
\times(\gbp-\grp)^2$ and we do not use $\gbp$ information for the faint stars with $\gbp>18$
which are affected by background underestimation \citep[see][]{evans2018, arenou2018}.
Moreover, we add quadratically $10$~mmag to the uncertainties to take into account the
systematics. We also take into account the 3~mmag/mag drift on the $G$ band \citep{arenou2018,
weiler2018}.

Concerning the Gaia astrometric information \citep{lindegren2018}, we use the filter described
in equation~1 of \cite{arenou2018} to remove astrometry with large $\chi^2$.  We correct the
parallax zero point of -0.03~mas \citep{lindegren2018,arenou2018} and remove
obvious outliers having $\varpi + 3 \times \sigma_{\varpi}< 0$. 

\section{FEDReD in a nutshell}
\label{sec:fedred}
In this work, we use the Field Extinction-Distance Relation Deconvolver (FEDReD) method,
presented in \cite{babusiaux2020}. We just sum up here the general process of FEDReD to analyse
a line of sight (LoS) and infer the relation between extinction and distance.

The algorithm works in two separated steps. The first one deals with the individual analysis of
each star contained in the LoS. It looks for $\probc{O_j}{\ao,D}$, i.e. the likelihood
of an observed star $O_j$ (considering its apparent magnitudes and possibly its parallax) to be at a distance $D$
with extinction $\ao$ (absorption at 550~nm). To compute this probability, FEDReD compare the apparent photometry
of the star to an empirical Hertzprung-Russel diagram based on Gaia DR2 representative of local thin disk stars. 
We take into account the colour and extinction dependant extinction coefficients by using \cite{danielski2018} 
models with the same coefficients as used in \cite{lallement2019}.

Once we have computed every stars' density $\probc{O_j}{\ao,D}$ in the LoS, we merge them to obtain
an estimate of
the joint distribution of extinction and distance of the
entire field of view, $\probhat{\ao,D}$, with a Bayesian 
iterative Richardson-Lucy deconvolution \citep{richardson1972, lucy1974}. In other words, the prior of
$\probind{\ao,D}{k}$ of the $k^{th}$ iteration is the posterior of the ${k-1}^{th}$ iteration.
To initiate the process, we build a simple prior $\probind{\ao,D}{0}$ by multiplying two 
prior distributions, the distance distribution of stars, $\probind{D}{0}$ and the distribution
of extinction given the distance, $\probcind{\ao}{D}{0}$. The prior $\probind{D}{0}$ just
follows a square law of the distance to take into account the cone effect. The prior
$\probcind{\ao}{D}{0}$ is null where $\ao > 10 \times D$ and flat elsewhere, this condition being
experimentally verified using the map of \cite{lallement2019}, see \cite{babusiaux2020} for more details.

As we take into account the completeness, the result of the deconvolution is actually
the estimate of the probability
distribution $\probchat{\ao,D}{S}$ where $S$ is the completeness of the LoS, which is here the completeness of the 2MASS photometry.
To model the completeness we estimate the probability distribution of $\probc{S}{\ao,D}$ using
the empirical HR diagram and a rough completeness model. This distribution is used both
during the deconvolution process to {obtain $\probchat{\ao,D}{S}$ and from it to derive
the final $\probhat{\ao,D}$ distribution.

To obtain the relation $\ao(D)$ from the previous distribution, we use a Monte-Carlo process
to draw monotonic increasing relations in the distribution of $\probhat{\ao,D}$. We assess the probability of each Monte-Carlo Solution
(MCS) and we keep the 1\,000 best MCSs, which correspond to the 1\,000 best relations $\ao(D)$ of
the LoS. Finally, we fit a constrained median cubic spline through the MCSs using \texttt{cobs} R library \citep{ng2007}, to get the best
fit for the given LoS.

We discretise the extinction - distance space that we probe. We choose here for the extinction a sampling from $0$ to $30~\magnitude$ with a step
of $0.05~\magnitude$. Concerning the distance space it covers $0.1$ to 30~kpc but the step
is linear in the distance modulus space with a width of $0.05~\magnitude$.

\section{Merging line of sight results}
\label{sec:merging}
To obtain an extinction map of the Galactic disk, we split it into small fields of view and
we analyse them separately with FEDReD. 
To ensure the continuity between LoS, half of each LoS is shared with its 2
neighbour fields. So in practice it means that each star is contributing to the information of two contiguous LoS.
This allows an efficient post-processing. Indeed, the
best fit output of FEDReD can be polluted by two major effects.
On one hand, the deconvolution can lead to very noisy $\prob{\ao,D}$, in particular in crowded fields where outliers can be numerous. 
On the other hand, several dust clumps could be present in the field of view and create an angular
differential extinction which is difficult to handle with a single median spline fit. Thus we use 
the MCS information coming from neighbours fields of view to remove outliers solutions and smooth the results. 

To do so, we remove every MCS which is not included in the envelope drawn by the maximum and
minimum values of the two neighbour MCS envelops. We repeat the operation until no more MCS is removed,
30 to 60 iterations being usually needed to clean every LoS. This step converges
thanks to the fact that two neighbour fields overlap, ensuring the consistency of two
consecutive fields of view.

Once our sample is cleaned we can draw a map by picking randomly one of the remaining MCS for
each LoS. We smooth this map by averaging each field with its two direct neighbours
applying weights $1/2$, $1/4$ and $1/4$ for centre field and the sides respectively,
as each LoS is shared with its 2 neighbour fields.
We draw 1\,000 of these smoothed maps. 
We then fit a median constrained cubic spline on solutions using \texttt{cobs} to get the $\ao(D)$ relation of each LoS.

To obtain the map of extinction density $a_0$ we decumulate the $\ao(D)$ relations by
processing the difference between subsequent distance bins and normalise by the
width of each distance bin.

Finally, we determine minimum and maximum valid distance
for each field of view. This distance interval is defined by the observability of a red clump
star in the infrared photometric band, considering the completeness of the field of view.
We
use red clump stars because their small intrinsic dispersion of absolute magnitude and colour leads to 
 individual distributions $\probc{O_j}{\ao,D}$ being more peaked
than other stellar types, so they bring more constraints \citep{babusiaux2020}.
We process the theoretical apparent magnitude of a red clump star by using the absolute
magnitudes inferred by \cite{ruiz-dern2018}
($M_J=-0.95~\magnitude$, $M_H=-1.45~\magnitude$ and $M_K=-1.61~\magnitude$) and the best MCS
for the $\ao(D)$ relation. Valid distances correspond to the ones where the red clump apparent
magnitude is between the saturation and the completeness magnitudes of the field. 

\begin{figure*}
  \centering
  \includegraphics[width=\linewidth]{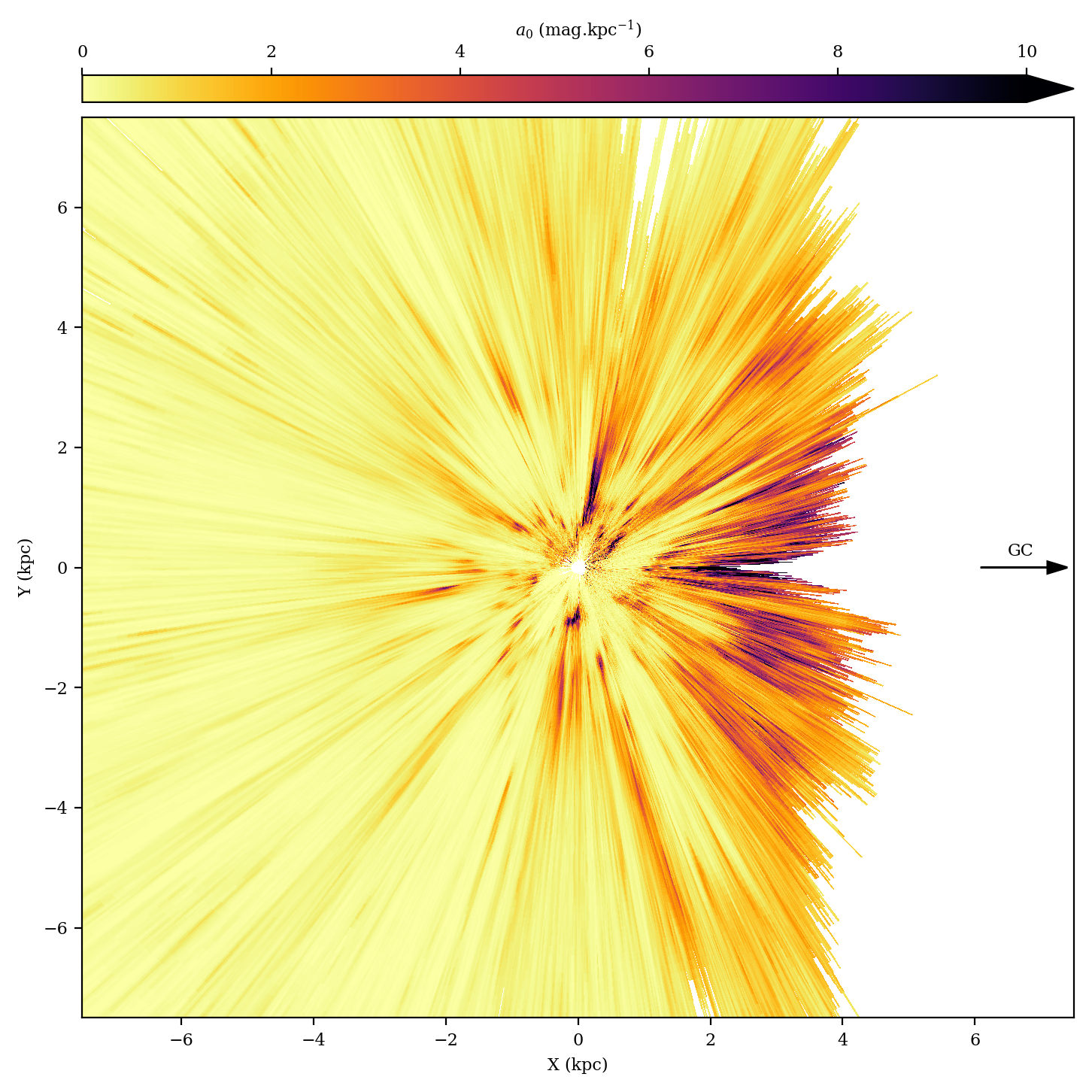}
  \caption{Density of extinction $a_0$ at $b=0\deg$. The Sun is at (0,0) and the Galactic centre direction
    is shown by the black arrow.
  }
  \label{fig:extMap}
\end{figure*}

\begin{figure}[ht]
  \centering
  \includegraphics{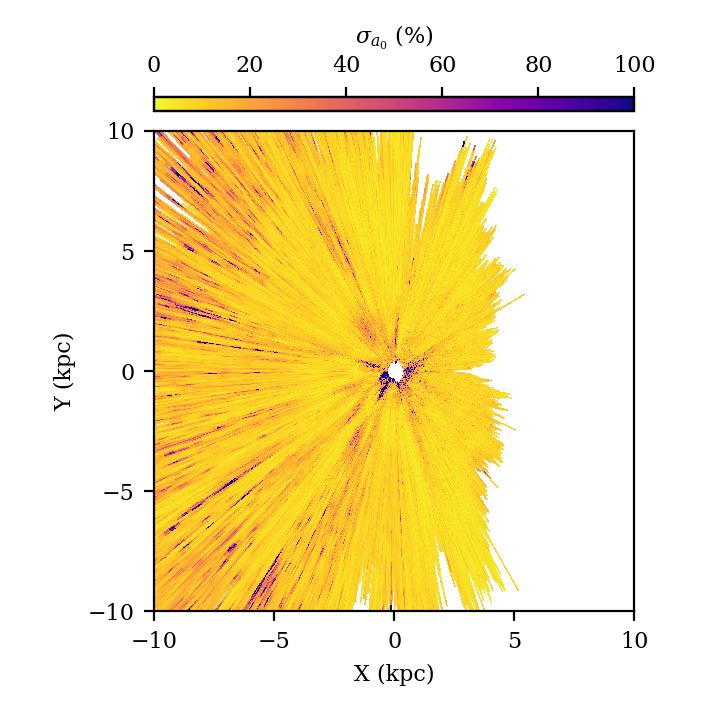}
  \caption{map of the relative error on extinction density, obtained by bootstraps.}
  \label{fig:errorMap}
\end{figure}

\section{Results}%
\label{sec:results}
The method described above has been applied to $3\,764$ LoS. Each LoS is centred at the
Galactic latitude $b=0\deg$ with a latitude width of $0.48\deg$.  The longitude width and
position are defined so as to obtain fields of view which contain 3\,000 stars and share
$1\,500$ stars with each neighbours. This overlap  with adjacent fields is necessary to
properly clean MCS by the neighbours minimum, maximum envelope. The average width of a LoS in
first and fourth quadrant is $0.12\deg$ whereas the average width is $0.43\deg$ in the second and
third quadrant. This difference is due to the larger number of observed stars in the central region
of the Galaxy.  We represent on Figure~\ref{fig:extMap} the resulting extinction density map in
the Galactic plane. The white area centred on the Sun corresponds to the too close distances where
the red clump is saturated (see previous section).

\subsection{Uncertainty on the extinction and on the extinction density}
\label{sub:error}
By using the clean sample of MCSs we can obtain the minimum and maximum cumulated
extinction at each distances (${\ao}_{\min}(D)$ and ${\ao}_{\max}(D)$). 

We also compute an error map of the extinction density. Basically,  we use our sample of MCSs
cleaned of outliers and we apply a bootstrap technique. To do so, we randomly keep a subsample of
MCSs for each line of sight and we apply the algorithm previously described to obtain a
bootstrapped map. We compute $100$ of these bootstrapped maps and then process the standard deviation
at each $(l,D)$ location to obtain the extinction density uncertainty.
Figure~\ref{fig:errorMap} presents the relative uncertainty.

This error map mostly represents the sampling error, so it underestimates the true uncertainty of
our results. The few locations where the relative uncertainty is really high correspond to
areas with almost no extinction or areas with a high-increasing rate of extinction density
(\textit{i.e.} a high $da_0/dD$) so
the algorithm can not locate exactly the beginning of the cloud.
We also see that the uncertainty increase at large distances towards the anticentre.

\subsection{Main Structures}%
On Figure~\ref{fig:labeledMap} we label the main features visible on our map. 
We also overplot some other ISM tracers on the map, in order to obtain a better view of the
ISM structures. In Figure~\ref{fig:molecCloud} we add the molecular clouds of
\cite{miville-deschenes2016}. Figure~\ref{fig:h2region} presents the \HII\ regions locations
inferred by \cite{hou2014} as well as the location of holes in the
 young stars distribution found by \cite{chen2019b}. Finally, Figure~\ref{fig:maser_map} represents
masers from \cite{reid2019}.

\begin{figure}[h]
  \centering
  \includegraphics{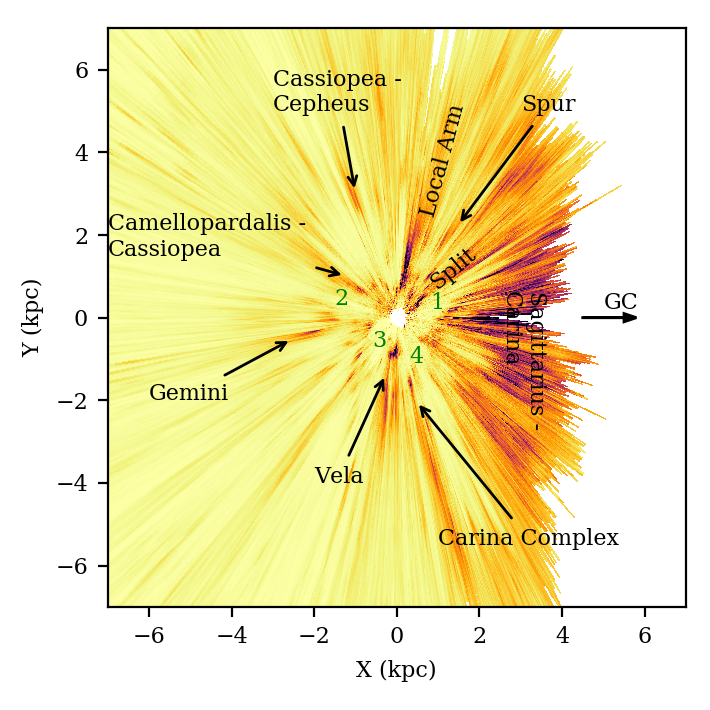}
  \caption{Extinction map with labelled structures. The green numbers correspond to the four
  bubbles which delineate an empty region around the Sun with a butterfly shape (see text for description).}
  \label{fig:labeledMap}
\end{figure}

\begin{figure}[ht]
  \centering
  \includegraphics{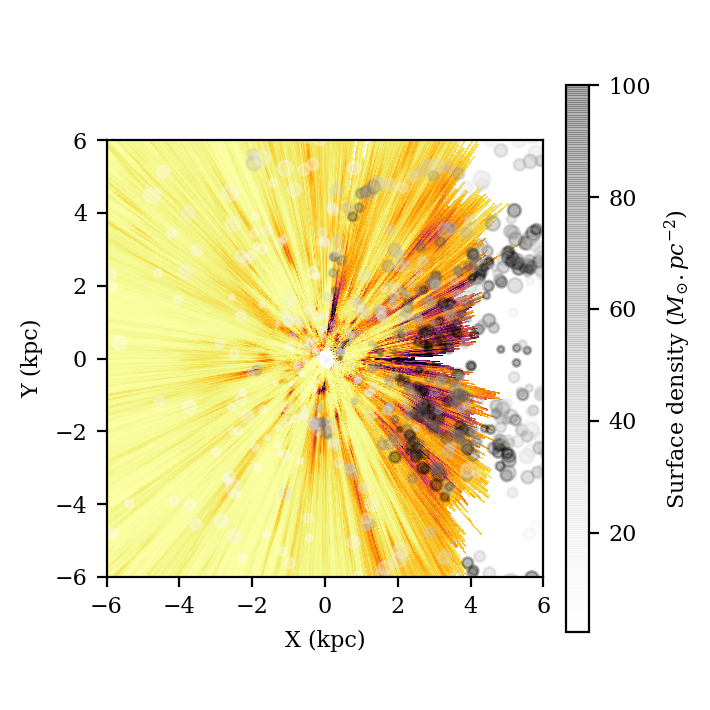}
  \caption{Extinction map with \cite{miville-deschenes2016} CO clouds within $|b|<0.24\deg$.
  Dot size is proportional to cloud size and the grey scale corresponds to the surface
density.}
  \label{fig:molecCloud}
\end{figure}

\begin{figure}[ht]
  \centering
  \includegraphics{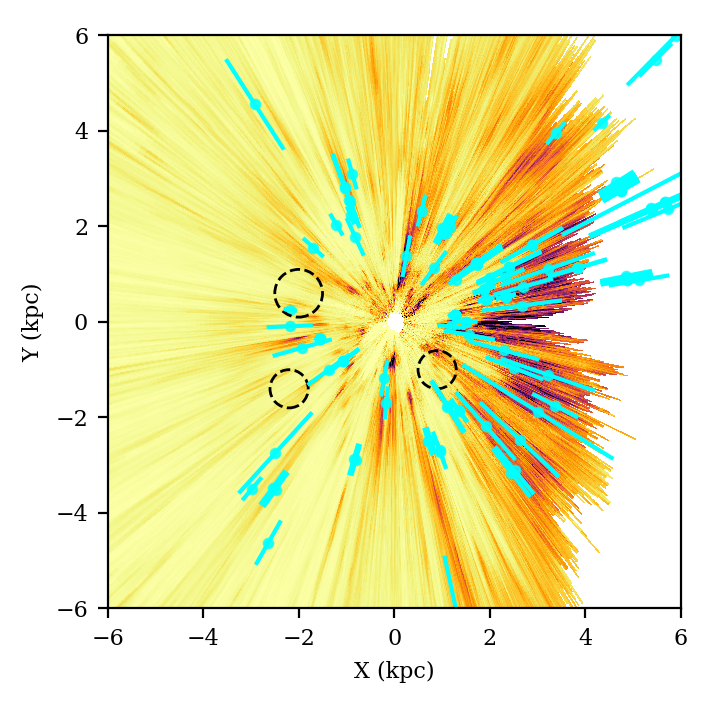}
  \caption{Extinction map with the \HII\ regions from \cite{hou2014}, keeping only regions
  with stellar distance information and with $|b|\leq0.24\deg$. Circles correspond to holes in
the young stars distribution found by \cite[Figure 4]{chen2019b}.}
  \label{fig:h2region}
\end{figure}

\begin{figure}[ht]
  \centering
  \includegraphics{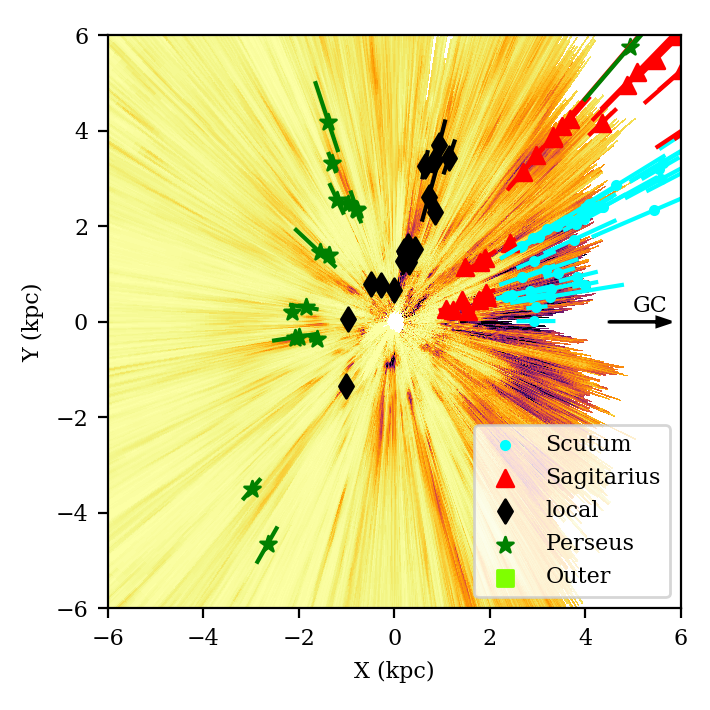}
  \caption{Extinction Map with masers from \cite{reid2019}. They are
    colour coded by their spiral arm membership. We only plot masers with $|z|<0.05~\kpc$.
  }
  \label{fig:maser_map}
\end{figure}

The biggest structure in our map is the Sagittarius-Carinae complex, which extends in a cone between
$\loneq{300}$ and $\loneq{30}$. Inside this high extinction area we can notice a cavity
centred on $\loneq{335}$, $D=2~\kpc$. This cavity presents an artefact which seems to be a \FoG.
This complex also contains a high density of molecular clouds and \HII\ regions except in the
cavity, which confirms the location and the size of this clean area.

Several high extinction structures related to spiral arms can be seen in the first quadrant.
The Local arm, at \loneq{80}, presents a very strong extinction consistent with the local arm masers of \cite{reid2019}.
The split \citep{lallement2019}, between $\loneq{30}$ and $\loneq{40}$, is almost parallel to the Local arm. 
We can also notice the extinction overdensity at $\loneq{60}$
between $D=1.5~\kpc$ and $D=3.5~\kpc$
which corresponds to the spur \citep{xu2018}.

In the second quadrant, two main extinction overdensities appear.
The first one is at \loneq{111} and \deq{3} and is associated with the Cassiopeia - Cepheus complex
\citep{ungerechts2000}. It coincides with the \cite{reid2019} masers associated with the Perseus
arm, and it is also well-marked by \HII\ regions.
The second overdensity is the Cameliopardalis-Cassiopeia cloud \citep{chen2014} which is at
\loneq{145} and \dbe{1}{3}.

The third quadrant contains many small high extinction areas associated with the local arm according
to the masers or to the Radcliffe wave \citep{alves2020}.
There is also the  Gemini molecular
cloud \citep{carpenter1995} at (\loneq{190}, \deq{2}) related to masers of the
Perseus arm.

The Vela cloud overlaps the third and the fourth quadrants. It exhibits a strong
overdensity in the foreground and is prolonged by an empty bubble surrounded by a thin
extinction edge. The distant boundary of this bubble is marked by molecular clouds and an HII region. 

Near the Vela cloud, at \loneq{282} and \deq{1.6}, the Carina complex \citep{zhang2001} also presents a
strong foreground structure and is prolonged up to \deq{6}. This large elongation is also
drawn by molecular clouds but the foreground structure only appears in extinction and is also
well visible in \cite{lallement2019}.

In addition to all of these structures, the Sun appears surrounded by four
bubbles without extinction, without taking into account the local bubble. The first one is
delimited by the Sagittarius-Carinae complex and the split. The second is centred on
$\loneq{165}$.The third one lies between the Local arm masers's
and the Vela cloud.
Finally, the longest one is surrounded by the Vela cloud and
Carina-complex on one side and by the Sagittarius-Carinae on the other side. Those four
extinction bubbles draw a structure that we will nickname ``the Butterfly''.
Those four bubbles also present a lack of every other ISM tracers. \cite{chen2019b} also note
holes in the distribution of young stars along the Perseus and Sagittarius arms, represented in Fig.~\ref{fig:h2region}. We notice in
\citet[Figure 4]{chen2019b} an under-density of young stars in each bubble of the butterfly,
except for the third one.

\subsection{Spiral Structure}%
\label{sub:spiral_structure}

\begin{figure}[ht]
  \centering
  \includegraphics{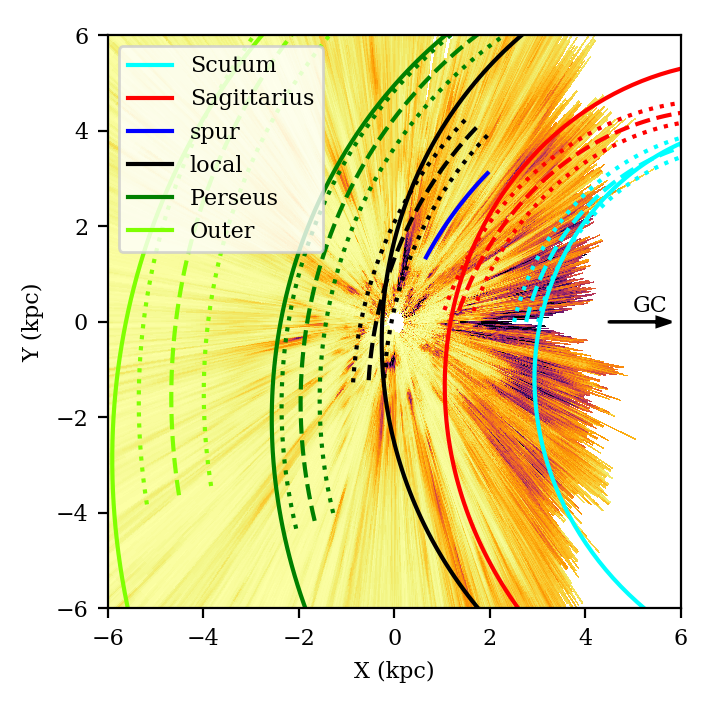}
  \caption{Extinction map with the spiral arms from \cite{reid2019} (dashed lines) and
  \cite{xu2018} for the spur and \cite{hou2014} (solid lines) overplotted.}
  \label{fig:spiralarm}
\end{figure}

On Figure~\ref{fig:spiralarm} we overplot the spiral arms fitted by \cite{reid2019} on masers
and by \cite{hou2014} on \HII\ regions. While most of the masers and \HII\ regions used to
create these models have a footprint on the dust map (Fig.\ref{fig:maser_map} and
\ref{fig:h2region}), the dust behaviour is very patchy and does not lead to obvious continuous
spiral arm footprints. 

The local arm is very well defined by the extinction in the first quadrant,
however, its path in the third/fourth quadrant is not obvious. According to masers, it seems that
the Local arm does not really reach the Sun and miss the Vela complex to follow the Radcliffe
wave \citep{alves2020}. On the other hand,  following the \cite{hou2014}
model, the Vela complex is part of the Local arm.

The Sagittarius arm shows a strong extinction but it is cut in two parts by the first clean
bubble. Moreover, it appears that the split of the Local arm, even if it seems to be
connected to the local arm at close distance, is more related to the Sagittarius arm at larger
distance. The spur looks like an extinction bridge between the Local and the Sagittarius
arms.

The Scutum arm crosses a high extinction area, however, we cannot distinguish a clear separation
between the Scutum and the Sagittarius extinction areas. They are merged in the Sagittarius-Carina complex.

The Perseus arm is mostly visible at masers's locations, it presents a very patchy structure
\citep{baba2018} and is only noticeable thanks to the visual guide of \cite{reid2014}.
Furthermore, the two masers of this arm at the bottom of the third quadrant are not even
visible in our extinction map because their extinction signature is too small for our spatial
resolution.

The outer arm is not visible in our extinction map. This is due to its high distance from the Galactic plane.

\subsection{Comparison with other work}%
\begin{figure}[h]
  \centering
  \includegraphics{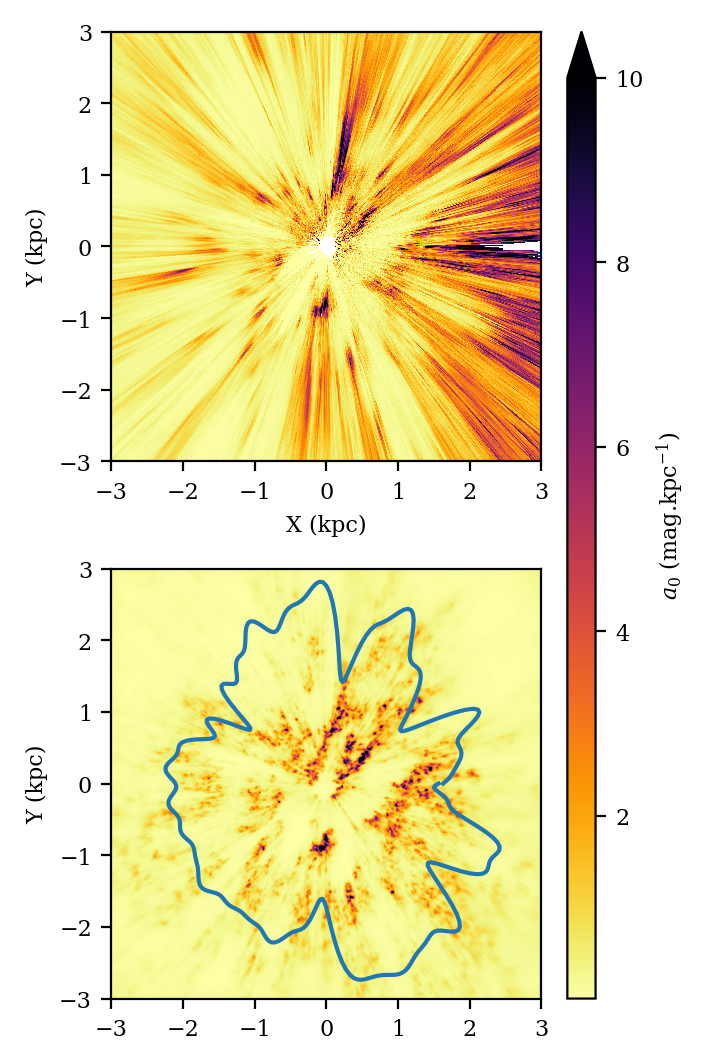}
  \caption{Comparison with \cite{lallement2019} results. Both maps are plotted with the same
  distance ranges and the same colour map. The blue curve represents the confidence limit of the
\cite{lallement2019} results.}
\label{fig:lallement_comp}
\end{figure}
We compared our extinction map to one of the most complete available extinction map of the
solar neighbourhood, by \cite{lallement2019}. They basically use the same data as our study (2MASS
and Gaia-DR2) limited to stars with a relative parallax error better than 20\%.  Their results
are presented in the reference extinction at $550 {\rm nm}$, as we do in this work. We present
in Figure~\ref{fig:lallement_comp} our map, truncated to the same distance range as
\cite{lallement2019} with the same colour map. 

The reader can notice the very good agreement between the two maps within the confidence
limit of \cite{lallement2019} results. The presence of {\FoG} on both maps 
is also apparent, though at different locations on the two maps, allowing us to really identify
them as spurious.

The foreground part of the Vela cloud is roughly identical on
both works, which confirms this crescent shape. This particular morphology tends to appear as
well on Fig. 12 of \cite{chen2019a}. 

The split of the Local arm, first described by \cite{lallement2019}, is also very strong in our
result. Its closest part is roughly the same as on the \cite{lallement2019} map. However, 
their method prevents them from detecting the elongation of the split as we do. This elongation
is less sloping than the foreground part, in agreement with \cite{chen2019a,green2019}.

The Cameliopardalis - Cassiopeia complex over density is at the same location on our work  and on
\cite{lallement2019}, \cite{green2019} and \cite{chen2019a}. 

However, we detect structures and prolongation of structures at larger distance which did not
appear in previous results in the literature.
For example, the complete elongation
of the Carina complex is too far to be visible on \cite{lallement2019} nor on \cite{chen2019a}.
In the same way the split of the local arm is only visible locally in \cite{lallement2019} which does
not permit to link it with the Sagittarius - Carina complex. \cite{chen2019a} and \cite{green2019} 
see a void between the split and Sagittarius but at a location which actually contains masers. 

\section{Conclusion}%
\label{sec:discussion}
We have used the Gaia DR2 data and 2MASS photometry to estimate the extinction within the Galactic plane. 
We used data of about 5.6 million stars to produce
an extinction map, reaching about $4~\kpc$ in the direction of the Galactic centre and more
than $6~\kpc$ in other directions. Thanks to this result we are able to confirm general
structures of the solar neighbourhood, already revealed by recent previous works
\citep{lallement2019, green2019, chen2019a}.
The Local arm and the Split represent two
structures, close to the Sun, with a high extinction separated by a clean corridor.
Nevertheless our map reveals a possible relation between the Split and the Sagittarius arm at
larger distance. The Galactic centre direction is dominated by the Sagittarius - Carina complex.
Because of this high extinction, as well as crowding, we are not able to
go beyond this complex. We do not distinguish the Sagittarius and the Scutum arm within this complex. The Perseus arm extinction component seems very fragmented, even if the
location of masers which trace this arm are visible in the dust. The Vela and Carina complex are the two strongest
extinction areas of the forth quadrant.
Furthermore, FEDReD's map reveals the extinction prolongation of the
Carina Complex and the bubble structure behind the Vela cloud. Finally, we also observe four empty bubbles close to the Sun, and we reveal
their ends for two of them. 

Data corresponding to the maps presented here are available in Tables~\ref{tab:results} and \ref{tab:errDens}.
We also provide the values of the cumulated extinction $\ao$ and corresponding asymmetric
uncertainties in Tables ~\ref{tab:Cumu}, \ref{tab:cumuMax} and \ref{tab:cumuMin}. The entire
version will be available as a machine-readable form at the CDS

In future works, we will also explore the Galactic disk at higher and lower latitudes. This will allow a better
study of some structures, for example the Outer arm which is known to be warped.

To explore larger distances, and in particular the Galactic centre and the start of spiral arms,
we will use deeper near-infrared surveys, such as UKIDSS \citep{lucas2008} and VISTA
\citep{minniti2010}. Moreover we look forward for the future third data release of Gaia mission which will provide better
parallax and photometry constraints.  

\begin{table*}
  \centering
  \caption{Some columns and rows of the $a_0$ extinction density (in mag/\kpc) of the map
    presented in Figure~\ref{fig:extMap}. Each column corresponds to a LoS, the longitude is given in the
    first row (in degree). The first column corresponds to the heliocentric distance in
    kiloparsecs. Blank spaces represent spatial
    locations outside of the distance confidence intervals (see
  section~\ref{sec:merging}). The full table is available in electronic form at the CDS.}
  \label{tab:results}
  \begin{tabular}{lccccc}
    \hline
    \hline
    &\multicolumn{5}{c}{$a_0$ ($\magnitude/\kpc$) at different longitude}\\
    \hline
    D (kpc) &  \loneq{0.11}   &  \loneq{0.23}   &  ...  &  \loneq{359.87} &
    \loneq{359.99} \\
    \hline
    0.11 &       &       &     & \\
    ...  & ...   & ...   & ... & ... \\
0.80     & 0.204 & 0.405 & ... &        & 0.153 \\
0.82     & 0.141 & 0.185 & ... & 0.048  & 0.100 \\
0.84     & 0.312 & 0.236 & ... & 0.096  & 0.171 \\
0.86     & 0.722 & 0.564 & ... & 0.290  & 0.367 \\
0.88     & 1.142 & 0.900 & ... & 0.489  & 0.569 \\
0.90     & 1.233 & 1.204 & ... & 0.680  & 0.649 \\
0.92     & 0.986 & 1.477 & ... & 0.863  & 0.604 \\
0.94     & 0.734 & 1.756 & ... & 1.050  & 0.559 \\
0.97     & 0.994 & 1.980 & ... & 1.289  & 0.713 \\
    ...         & ...   & ...   & ... & ... & ... \\
    \hline                 
  \end{tabular}           
\end{table*}              

\begin{table*}
  \centering
  \caption{Some columns and rows of the $A_0$ extinction (mag).  Each column corresponds to a
    LoS, the longitude is the first row (in degree). The first column corresponds to the
    heliocentric distance in kiloparsecs. Blank spaces represent spatial
    locations outside of the distance confidence intervals (see section~\ref{sec:merging}). The full table is available in electronic form at the CDS.}
  \label{tab:Cumu}
  \begin{tabular}{lccccc}
    \hline
    \hline
    &\multicolumn{5}{c}{$A_0$ ($\magnitude$) at different longitude}\\
    \hline
    D (kpc) &  \loneq{0.11}   &  \loneq{0.23}   &  ...  &  \loneq{359.87} &
    \loneq{359.99} \\
    \hline
    0.11 &       &       & ... &       & \\
    ...  & ...   & ...   & ... & ...   & ... \\
0.81     & 0.681 & 0.641 & ... & 0.775 & 0.764 \\
0.83     & 0.688 & 0.646 & ... & 0.777 & 0.768 \\
0.85     & 0.702 & 0.657 & ... & 0.783 & 0.775 \\
0.87     & 0.725 & 0.675 & ... & 0.793 & 0.787 \\
0.89     & 0.751 & 0.700 & ... & 0.807 & 0.800 \\
0.91     & 0.772 & 0.731 & ... & 0.825 & 0.813 \\
0.93     & 0.787 & 0.770 & ... & 0.848 & 0.825 \\
0.95     & 0.810 & 0.814 & ... & 0.876 & 0.841 \\
    ...         & ...   & ...   & ... & ... & ... \\
    \hline                  
  \end{tabular}
\end{table*}

\begin{table*}
  \centering
  \caption{Some columns and rows of the $a_0$ density uncertainty $\sigma_{a_0}$ (in
    $\magnitude/\kpc$) presented in  Figure~\ref{fig:errorMap} (see section~\ref{sub:error}).  Each
    column corresponds to a LoS, the longitude is the first row (in degree). The first column
    corresponds to the heliocentric distance in kiloparsecs. Blank spaces represent spatial
    locations outside of the distance confidence intervals. The full table is available in electronic form at the CDS.}
  \label{tab:errDens}
  \begin{tabular}{lccccc}
    \hline
    \hline
    &\multicolumn{5}{c}{$\sigma_{a_0}$ ($\magnitude/\kpc$) at different longitude}\\
    \hline
    D (kpc) &  \loneq{0.11}   &  \loneq{0.23}   &  ...  &  \loneq{359.87} &
    \loneq{359.99} \\
    \hline
    0.11 &       &       & ... &     & \\
    ...  & ...   & ...   & ... & ... & ... \\
0.80 &   0.111 &   0.109 &  ... &        &   0.055 \\
0.82 &   0.137 &   0.092 &  ... &  0.062 &   0.065 \\
0.84 &   0.145 &   0.104 &  ... &  0.067 &   0.086 \\
0.86 &   0.160 &   0.100 &  ... &  0.077 &   0.101 \\
0.88 &   0.199 &   0.077 &  ... &  0.062 &   0.108 \\
0.90 &   0.232 &   0.085 &  ... &  0.060 &   0.115 \\
0.92 &   0.221 &   0.079 &  ... &  0.090 &   0.128 \\
0.94 &   0.209 &   0.082 &  ... &  0.115 &   0.147 \\
0.97 &   0.307 &   0.108 &  ... &  0.150 &   0.170 \\
    ...         & ...   & ...   & ... & ... & ... \\
    \hline                    
  \end{tabular}
\end{table*}

\begin{table*}
  \centering
  \caption{Some columns and rows of the maximum values of the extinction $A_0$ (see section~\ref{sub:error}).  Each column
    corresponds to a LoS, the longitude is the first row (in degree). The first column
    corresponds to the heliocentric distance in kiloparsecs. Blank spaces represent spatial
    locations outside of the distance confidence intervals. The full table is available in electronic form at the CDS.}
  \label{tab:cumuMax}
  \begin{tabular}{lccccc}
    \hline
    \hline
    &\multicolumn{5}{c}{${A_0}_{\rm max}$ ($\magnitude$) at different longitude}\\
    \hline
    D (kpc) &  \loneq{0.11}   &  \loneq{0.23}   &  ...  &  \loneq{359.87} &
    \loneq{359.99} \\
    \hline
    0.11 &       &       & ... &     & \\
    ...  & ...   & ...   & ... & ... & ... \\
0.81 &   0.750 &   0.850 &  ... &  0.900 &   0.900 \\
0.83 &   0.750 &   0.850 &  ... &  0.900 &   0.900 \\
0.85 &   0.900 &   0.850 &  ... &  0.900 &   1.000 \\
0.87 &   0.900 &   0.850 &  ... &  0.900 &   1.000 \\
0.89 &   0.900 &   0.850 &  ... &  0.900 &   1.000 \\
0.91 &   0.900 &   0.850 &  ... &  0.900 &   1.000 \\
0.93 &   0.900 &   0.850 &  ... &  0.900 &   1.000 \\
0.95 &   1.000 &   1.000 &  ... &  0.900 &   1.000 \\
    ...         & ...   & ...   & ... & ... & ... \\
    \hline                      
  \end{tabular}
\end{table*}

\begin{table*}
  \centering
  \caption{Some columns and rows of the minimum values of extinction $A_0$  (see section~\ref{sub:error}).  Each column
    corresponds to a LoS, the longitude is the first row (in degree). The first column
    corresponds to the heliocentric distance in kiloparsecs. Blank spaces represent spatial
    locations outside of the distance confidence intervals. The full table is available in electronic form at the CDS.}
  \label{tab:cumuMin}
  \begin{tabular}{lccccc}
    \hline
    \hline
    &\multicolumn{5}{c}{${A_0}_{\rm min}$ ($\magnitude$) at different longitude}\\
    \hline
    D (kpc) &  \loneq{0.11}   &  \loneq{0.23}   &  ...  &  \loneq{359.87} &
    \loneq{359.99} \\
    \hline
    0.11        &       &       & ... &     & \\
    ...         & ...   & ...   & ... & ... & ... \\
0.81 &   0.550 &   0.550 &   ... &0.600 &   0.600 \\
0.83 &   0.600 &   0.550 &   ... &0.600 &   0.600 \\
0.85 &   0.600 &   0.550 &   ... &0.600 &   0.600 \\
0.87 &   0.600 &   0.550 &   ... &0.600 &   0.600 \\
0.89 &   0.600 &   0.600 &   ... &0.600 &   0.600 \\
0.91 &   0.600 &   0.600 &   ... &0.650 &   0.700 \\
0.93 &   0.600 &   0.700 &   ... &0.650 &   0.700 \\
0.95 &   0.650 &   0.750 &   ... &0.650 &   0.700 \\
    ...         & ...   & ...   & ... & ... & ... \\
    \hline                       
  \end{tabular}
\end{table*}

\begin{acknowledgements}
  We thank the referee for suggestions that improved the paper.
  This work has made use of data from the European Space Agency (ESA) mission
  {\it Gaia} (\url{https://www.cosmos.esa.int/gaia}), processed by the {\it Gaia}
  Data Processing and Analysis Consortium (DPAC,
  \url{https://www.cosmos.esa.int/web/gaia/dpac/consortium}). Funding for the DPAC
  has been provided by national institutions, in particular the institutions
  participating in the {\it Gaia} Multilateral Agreement.
  This publication also make use of data products from the Two Micron All Sky Survey, 
  which is a joint project of the University of Massachusetts and the Infrared Processing 
  and Analysis Center/California Institute of Technology, funded by the National Aeronautics 
  and Space Administration and the National Science Foundation.
  We benefitted from the computing resources of MesoPSL financed by the Region Ile de France and the 
  project Equip@Meso (reference ANR-10-EQPX-29-01).
\end{acknowledgements}


\bibliographystyle{aa} 
\bibliography{2MASS_extmap}

\end{document}